\documentclass[prd,aps,twocolumn,10pt,letterpaper,superscriptaddress]{revtex4-2}
\usepackage[utf8]{inputenc}
\usepackage[english]{babel}
\usepackage{graphicx}
\usepackage{float}
\usepackage{amsmath}
\usepackage{bm}
\begin{document}

\title{The influence of electron–electron interaction on pair production in supercritical collisions of highly charged ions}

\author{N. K. Dulaev}
\email{st069071@student.spbu.ru}
\affiliation{Department of Physics,
St.~Petersburg State University, 7-9 Universitetskaya nab., St.~Petersburg 
199034, Russia}
\affiliation{Petersburg Nuclear Physics Institute named by B.~P.~Konstantinov of National Research Center ''Kurchatov Institute'', Orlova roscha 1, 188300 Gatchina, Leningrad region$,$ Russia}

\author{D. A. Telnov}
\email{d.telnov@spbu.ru}
\affiliation{Department of Physics,
St.~Petersburg State University, 7-9 Universitetskaya nab., St.~Petersburg 
199034, Russia}

\author{R. V. Popov}
\affiliation{Petersburg Nuclear Physics Institute named by B.~P.~Konstantinov of National Research Center ''Kurchatov Institute'', Orlova roscha 1, 188300 Gatchina, Leningrad region$,$ Russia}
\affiliation{Department of Physics,
St.~Petersburg State University, 7-9 Universitetskaya nab., St.~Petersburg 
199034, Russia}

\author{V. M. Shabaev}
\email{v.shabaev@spbu.ru}
\affiliation{Department of Physics,
St.~Petersburg State University, 7-9 Universitetskaya nab., St.~Petersburg 
199034, Russia}
\affiliation{Petersburg Nuclear Physics Institute named by B.~P.~Konstantinov of National Research Center ''Kurchatov Institute'', Orlova roscha 1, 188300 Gatchina, Leningrad region$,$ Russia}

\author{Y.~S.~Kozhedub}
\affiliation{Department of Physics,
St.~Petersburg State University, 7-9 Universitetskaya nab., St.~Petersburg 
199034, Russia}

\author{X.~Ma}
\affiliation{Institute of Modern Physics,
Chinese Academy of Sciences, 730000 Lanzhou, China}

\author{I. A. Maltsev}
\affiliation{Department of Physics,
St.~Petersburg State University, 7-9 Universitetskaya nab., St.~Petersburg 
199034, Russia}

\author{A. D. Mironov}
\affiliation{Department of Physics,
St.~Petersburg State University, 7-9 Universitetskaya nab., St.~Petersburg 
199034, Russia}

\author{I. I. Tupitsyn}
\affiliation{Department of Physics,
St.~Petersburg State University, 7-9 Universitetskaya nab., St.~Petersburg 
199034, Russia}

\begin{abstract}
The effect of electron-electron interaction on positron emission in supercritical collisions of highly charged ions is studied within the monopole approximation using the time-dependent density functional theory and the time-dependent Hartree-Fock-Slater methods. Positron production probabilities and energy spectra are calculated for U–U, U–Cm, and Cm–Cm collision systems, considering both bare nuclei and highly charged ions with partially filled electron shells. The results demonstrate that electron screening substantially reduces positron production and suppresses the characteristic signatures of spontaneous vacuum decay, previously found in collisions of bare nuclei.
\end{abstract}

\maketitle

\section{Introduction}

Since the development of quantum electrodynamics (QED), considerable attention has been devoted to phenomena occurring in the non-perturbative regime of extremely strong electromagnetic fields. QED predicts that when the strength of a static uniform electric field exceeds a critical threshold~--- on the order of $10^{16}$ V/cm~--- the vacuum becomes unstable and can spontaneously create electron-positron pairs. The experimental observation of this process would provide a direct test of QED in the supercritical regime.

Despite significant advances in high-intensity laser technology (see, e.g., Refs. \cite{Fedotov_2023_Advances, Kostyukov_2024_Physics}), the field strengths achieved in laboratory conditions remain several orders of magnitude below the supercritical threshold. An alternative route to probing this regime is provided by the Coulomb field of superheavy nuclei with charge numbers exceeding a critical value, $Z>Z_{\textrm{cr}}$, where $Z_{\textrm{cr}}\approx 173$. However, no such nuclei have been produced yet. Instead, heavy-ion collisions offer a means to achieve supercritical fields, making them a subject of theoretical interest for more than half a century. Early studies by Soviet and German physicists \cite{Gershtein_1969, Pieper_1969_Interior, Popov_1970_1, *Popov_1970_2, *Popov_1970_3, *Popov_1971_1, Zeldovich_1971_Electronic, Muller_1972_Solution, *Muller_1972_Electron, Mur_1976, Popov_1976, Muller_1976_Positron, Reinhardt_1977_Quantum, Soff_1977_Shakeoff, Rafelski_1978_Fermions, Greiner_1985_Quantum} demonstrated that in collisions where the combined nuclear charge exceeds $Z_{\textrm{cr}}$ the resulting Coulomb field can reach supercritical strength at sufficiently small internuclear distances $R_{\textrm{cr}}$. Under these conditions, the lowest unoccupied quasimolecular electronic state, $1s\sigma_{1/2}$, may dive into the negative-energy continuum, leading to the spontaneous creation of electron-positron pairs. The emitted positrons can escape and be detected, while the corresponding electrons remain bound, creating the charged vacuum.

However, direct observation of spontaneous vacuum decay in heavy-ion collisions faces significant challenges. In slow collisions ($\sim 0.1c$), the supercritical field persists for only $10^{-21}$ s, which is two orders of magnitude shorter than the spontaneous resonance lifetime, leading to a low probability of spontaneous pair production. Furthermore, spontaneous pair creation is strongly masked by the dynamical pair production induced by the time-dependent Coulomb field of the moving nuclei. These two mechanisms interfere coherently and cannot be easily distinguished. Thus, theoretical investigations by the Frankfurt group  (see, e.g., \cite{Smith_1974_Induced, Reinhardt_1981_Theory, Muller_1988_Positron, Rafelski_1978_Fermions, Greiner_1985_Quantum, Bosch_1986_Positron, Muller_1994_Electron, Reinhardt_2005_Supercritical, Rafelski_2017_Probing}), suggested that spontaneous pair production could be observable only if the colliding nuclei form a quasi-bound state (``nuclear sticking''), prolonging the supercritical phase and enhancing the spontaneous channel. However, no experimental evidence for such sticking behavior with heavy ions has been found to date. This motivates the search for alternative approaches to detect vacuum decay in supercritical fields.

Over the past decade, the theoretical research group at Saint Petersburg State University has made significant progress in identifying possible experimental approaches for observing spontaneous vacuum decay in heavy-ion collisions. In Refs. \cite{Maltsev_2019_How, Popov_2020_How}, within the framework of the monopole approximation, it was demonstrated that a transition to the supercritical regime can be detected in collisions where nuclei follow trajectories with a fixed minimal internuclear distance $R_{\mathrm{min}}$ and varying energy parameter $\varepsilon=E/E_0$, where $E$ is the collision energy and $E_0$ is the energy of the head-on collision. In this approach in supercritical collisions as $\varepsilon$ decreases, the probability of pair production increases while the contribution of the dynamical mechanism diminishes due to the decreasing collision velocity. Further studies beyond the monopole approximation \cite{Popov_2023_Spontaneous} confirmed the signatures of the spontaneous pair production regime. Moreover, recent two-center calculations \cite{Dulaev_2024_Angular, Dulaev_Three_2025} have enabled an analysis of the angular distributions of emitted positrons, taking into account rotational coupling effects.

With the upcoming experimental facilities in Germany (GSI/FAIR) \cite{Gumberidze_2009_X, Lestinsky_2016_Physics}, China (HIAF) \cite{Ma_2017_HIAF}, and Russia (NICA) \cite{Akopian_2015_Layout}, the study of pair production in heavy-ion collisions is becoming increasingly relevant. One of the key challenges in experimentally implementing the approach proposed in \cite{Maltsev_2019_How, Popov_2020_How} is the production of low-energy beams of fully stripped ions. In practice, partially stripped ions are more commonly available. The Frankfurt research group conducted extensive studies on electron and positron emission in subcritical and supercritical collision systems with filled lowest electron shells \cite{Reus_1984_The, Muller_1984_Electron, Reus_1987_Positron, Muller_1988_Positron, Reus_2022_Positron}. Their results indicate that in collisions involving partially ionized atoms, pair production probabilities are significantly suppressed due to Pauli blocking, with electron screening providing an additional reduction in the total probability.

In this work, we investigate the effect of electron-electron interaction on the signatures of spontaneous positron production in heavy-ion collisions within the supercritical regime. Calculations are performed for U–U, U–Cm and Cm–Cm collision systems within a monopole approximation. The collisions of bare nuclei and of partially stripped ions with lowest filled shells are considered. The electron-electron interaction is included within two distinct mean-field approaches. The total positron creation probabilities as well as positron spectra are calculated and discussed. 

Atomic units ($\hbar=|e|=m_{e}=1$) are used throughout the paper unless specified otherwise. 

\section{Methods}\label{methods}
In our method, a collision of two heavy ions is described within the monopole approximation, which proved to provide an adequate description of the electron-positron pair production \cite{Popov_2023_Spontaneous}. The electron-electron interaction in the initially occupied orbitals is treated using a mean-field approach, so we solve a set of one-electron time-dependent Dirac equations (TDDE) for independent electrons initially occupying the negative energy Dirac continuum and a number of bound-state orbitals (up to the Fermi level $F$):
\begin{equation}
    i \frac{\partial}{\partial t} \Psi(\bm{r}, t) = H\Psi(\bm{r},t), \label{eq:tdde}
\end{equation}
where $\Psi(\bm{r},t)$ is a four-component one-electron orbital, and the Hamiltonian $H$ can be written as
\begin{equation}
    H = c(\bm{\alpha} \cdot \bm{p}) + c^2 \beta + V_{\mathrm{n}}(r, t) + V_{\mathrm{MF}}(r, t).
    \label{eq:hamiltonian}
\end{equation}
Here, $c$ is the speed of light, $\bm{p}$ is the momentum operator, $\bm{\alpha}$ and $\beta$ are the Dirac matrices.

The spherically symmetric potential $V_{\mathrm{n}}(r, t)$ in Eq.~(\ref{eq:hamiltonian}) represents the interaction with the nuclei in the monopole approximation, and its dependence on time is due to variation of the internuclear distance.  The spherically symmetric mean-field potential $V_{\mathrm{MF}}(r, t)$ is produced by the electrons in the initially occupied bound states. This potential is constructed by a self-consistent procedure before solving the time-dependent equation (\ref{eq:tdde}), and its time dependence is due to the time evolution of the one-electron orbitals.

To represent the mean-field potential $V_{\mathrm{MF}}(r, t)$ and solve the TDDE (\ref{eq:tdde}), we make use of two different approaches. The first approach is
based on the time-dependent density-functional theory (TDDFT). The mean-field potential $V_{\mathrm{MF}}(r, t)$ comprises the Hartree electron-electron repulsion potential, the exchange-correlation potential in the local-density approximation, and the Fermi--Amaldi self-interaction correction. Eq.~(\ref{eq:tdde}) is solved numerically with the help of generalized pseudospectral method, which has been extensively used and discussed in earlier works \cite{Telnov_2007_Ab, Telnov_2009_Effects, Telnov_2018_Multiphoton}. 
The second approach employs a basis-set representation of the wave function in the TDDE (\ref{eq:tdde}) using $B$-splines, as it was implemented in the earlier work \cite{Popov_2020_How}. In this case, the mean-field potential $V_{\textrm{MF}}(r, t)$ is calculated using the time-dependent Hartree--Fock--Slater (TDHFS) method, following the methodology developed by the Frankfurt group \cite{Reus_1984_The}. In both schemes described above, the Crank-Nicolson algorithm \cite{Crank_1947_A} is applied to perform the time evolution. Once the ions collide and move apart, the positron creation probabilities and energy spectra are calculated by analyzing the hole distributions in the negative-energy Dirac continuum.
The pair-creation probabilities and positron spectra are calculated according to the methods presented in Refs. \cite{Popov_2020_How, Dulaev_2024_Angular}.

\section{Results}\label{results}

The total positron creation probabilities and energy distributions have been computed within the monopole approximation for U–U, U–Cm, and Cm–Cm collisions. The nuclear trajectories were modeled as classical Rutherford trajectories. Calculations were performed for both bare nuclei (Fermi level $F=0$) and partially stripped ions with the filled $1s_{1/2}$, $2s_{1/2}$, $3s_{1/2}$, $2p_{1/2}$, $3p_{1/2}$, and $4p_{1/2}$ orbitals ($F=3$). The case of $F=3$ was extensively studied by the Frankfurt group \cite{Reus_1984_The, Muller_1984_Electron, Reus_1987_Positron, Muller_1988_Positron, Reus_2022_Positron} as it closely models the GSI experiments \cite{Greiner_1985_Quantum}. The numerical computations were performed on a pseudospectral radial grid consisting of 1600 points with 16384 time propagation steps. The spherically symmetric nuclear potential of the monopole approximation was obtained by averaging the two-center nuclear potential on a $80$-point angular Gauss--Lobatto grid.

\subsection{Total positron creation probabilities}

To validate the reliability of the employed methods, we calculated the total positron creation probabilities for U–Cm and Cm–Cm collisions with $F=3$ at collision energies of $E\approx 748$~MeV and $E\approx 781$~MeV, respectively, including electron-electron interaction. Tables \ref{table:u-cm} and \ref{table:cm-cm} compare the results obtained in this work with those from Ref.~\cite{Reus_2022_Positron}, where calculations were performed using the adiabatic TDHFS method. As shown in the tables, the computed probabilities exhibit good agreement with previous results.
\begin{table}[]
\centering
\setlength\tabcolsep{3.5pt}
\caption{Total positron creation probabilities for U–Cm collisions at $E\approx 748$~MeV with $F=3$. Columns 2 and 3 present the results obtained in this work, while column 4 provides the corresponding values from Ref. \cite{Reus_2022_Positron} for comparison.}
\begin{tabular}{cccc} \hline\hline
$b$, fm  & TDDFT approach   & TDHFS approach       & Ref. \cite{Reus_2022_Positron}               \\ \hline
0  & 5.74$\times 10^{-4}$      & 5.59$\times 10^{-4}$     & 6.21$\times 10^{-4}$ \\
10 & 2.78$\times 10^{-4}$      & 2.70$\times 10^{-4}$     & 2.99$\times 10^{-4}$ \\
20 & 7.01$\times 10^{-5}$      & 6.78$\times 10^{-5}$     & 7.35$\times 10^{-5}$ \\
30 & 1.62$\times 10^{-5}$      & 1.55$\times 10^{-5}$     & 1.73$\times 10^{-5}$ \\ \hline\hline
\end{tabular} \label{table:u-cm}
\end{table}

\begin{table}[]
\centering
\setlength\tabcolsep{3.5pt}
\caption{Total positron creation probabilities for Cm–Cm collisions at $E\approx 781$~MeV with $F=3$. Columns 2 and 3 present the results obtained in this work, while column 4 provides the corresponding values from Ref. \cite{Reus_2022_Positron} for comparison.}
\begin{tabular}{cccc} \hline\hline
$b$, fm  & TDDFT approach   & TDHFS approach        & Ref. \cite{Reus_2022_Positron}        \\ \hline
0  & 8.91$\times 10^{-4}$      & 8.77$\times 10^{-4}$ & 8.92$\times 10^{-4}$ \\
10 & 4.30$\times 10^{-4}$      & 4.23$\times 10^{-4}$ & 4.37$\times 10^{-4}$ \\
20 & 1.09$\times 10^{-4}$      & 1.06$\times 10^{-4}$ & 1.11$\times 10^{-4}$ \\
30 & 2.52$\times 10^{-5}$      & 2.45$\times 10^{-5}$ & 2.67$\times 10^{-5}$ \\ \hline\hline
\end{tabular} \label{table:cm-cm}
\end{table}

To investigate signatures of the spontaneous pair production regime, we computed the total positron creation probabilities for U–U and Cm–Cm collisions at $\varepsilon=1.0$, $1.02$ and $1.05$ with a minimum internuclear distance of $R_{\mathrm{min}}=17.5$~fm. The results are presented in Table \ref{table:total_probs_f=3}, where column $3$ lists the probabilities for collisions of bare nuclei, while columns $4$ and $5$ provide the corresponding values for the ion collisions with the $1s_{1/2}-3s_{1/2}$ and $2p_{1/2}-4p_{1/2}$ shells filled ($F=3$), both with and without the screening effect due to the electron-electron interaction.

From the bare nuclei results, it is evident that for $Z=92$ the probability remains nearly constant with increasing $\varepsilon$, whereas for $Z=96$, it decreases. This behavior is consistent with the supercritical regime previously discussed in Refs.~\cite{Maltsev_2019_How, Popov_2020_How, Dulaev_Three_2025}. When the $1s_{1/2}-3s_{1/2}$ and $2p_{1/2}-4p_{1/2}$ shells are occupied ($F=3$, column $4$), the absolute probability is reduced by $1–2$ orders of magnitude. For both $Z=92$ and $Z=96$ collisions, the characteristic signatures of the supercritical regime observed for bare nuclei disappear, as the total probability now increases with $\varepsilon$, contrary to the bare-nuclei case. 

This behavior is readily understood: in collisions of bare nuclei, the $1s_{1/2}$ state serves as the primary destination for the electrons created by the spontaneous pair-production mechanism, when this state dives into the negative-energy continuum in the course of collision.  However, in the $F=3$ configuration, this state is initially occupied. If it remains occupied throughout the collision,  the related channel of positron emission is closed because of the Pauli blocking. Of course, a vacancy can be created in the $1s_{1/2}$ state during the collision thus opening the spontaneous positron emission channel but the probability of such a process appears quite small. 
As a result, in the $F=3$ case positron creation is dominated by transitions of electrons from the negative-energy continuum to the higher-lying bound states, which never dive into the negative-energy continuum during the collision; therefore, the signatures of the spontaneous positron creation mechanism are not observed. Accounting for the electron-electron interaction through the mean-field potential further reduces the total positron-creation probabilities without altering the overall trend.

\begin{table}[]
\centering
\setlength\tabcolsep{3.5pt}
\caption{Total positron creation probabilities for U–U and Cm–Cm collisions at $R_{\mathrm{min}}=17.5$~fm for different Fermi level configurations. Column $3$ lists the results for bare nuclei, while columns $4$ and $5$ contain the probabilities for ions with $F=3$, without and with electron-electron interaction, respectively.}
\begin{tabular}{ccccc} \hline\hline
$Z$                   & $\varepsilon$  & $F=0$      & $F=3$ & $F=3$  \\
&&& no screening&with screening\\ \hline
92 & 1.00    & 1.12$\times 10^{-2}$ & 3.63$\times 10^{-4}$   & 2.98$\times 10^{-4}$ \\
                    & 1.02 & 1.12$\times 10^{-2}$ & 3.85$\times 10^{-4}$   & 3.17$\times 10^{-4}$ \\
                    & 1.05 & 1.12$\times 10^{-2}$ & 4.19$\times 10^{-4}$   & 3.48$\times 10^{-4}$ \\ \hline
96 & 1.00    & 3.87$\times 10^{-2}$ & 9.24$\times 10^{-4}$   & 7.62$\times 10^{-4}$ \\
                    & 1.02 & 3.84$\times 10^{-2}$ & 9.74$\times 10^{-4}$   & 8.05$\times 10^{-4}$ \\
                    & 1.05 & 3.79$\times 10^{-2}$ & 1.05$\times 10^{-3}$   & 8.72$\times 10^{-4}$ \\  \hline\hline
\end{tabular}\label{table:total_probs_f=3}
\end{table}

\subsection{Energy distributions of positrons}
\begin{figure}[t]
    \centering
    \includegraphics[width=\columnwidth]{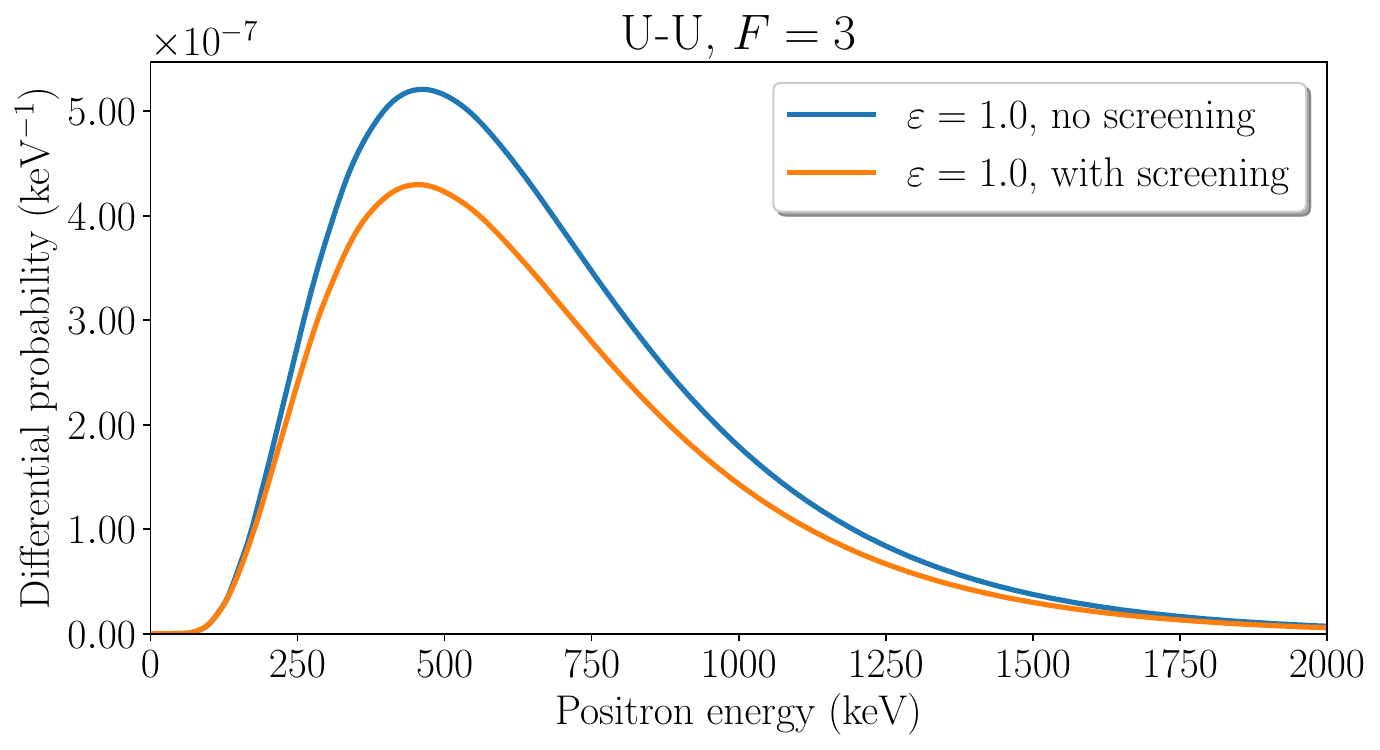}
    \caption{Energy spectra of positrons for symmetric head-on collision of ions with $Z=92$ and $F=3$ at $R_{\mathrm{min}}=17.5$~fm with and without electron-electron interaction.}
    \label{fig:z_92_F_3}
\end{figure}

Our findings about signatures of the supercritical positron creation regime are further supported by the calculations of the energy distributions of outgoing positrons. Fig. \ref{fig:z_92_F_3} presents the positron energy distributions for head-on U–U collisions with $R_{\mathrm{min}}=17.5$~fm at the Fermi level $F=3$. The results are shown both with and without accounting for the electron-electron interaction. From the figure it is evident that the inclusion of electron-electron interaction in the calculations leads to an overall reduction of the spectra while leaving their qualitative behavior nearly unchanged.
\begin{figure}
    \centering
    \includegraphics[width=\columnwidth]{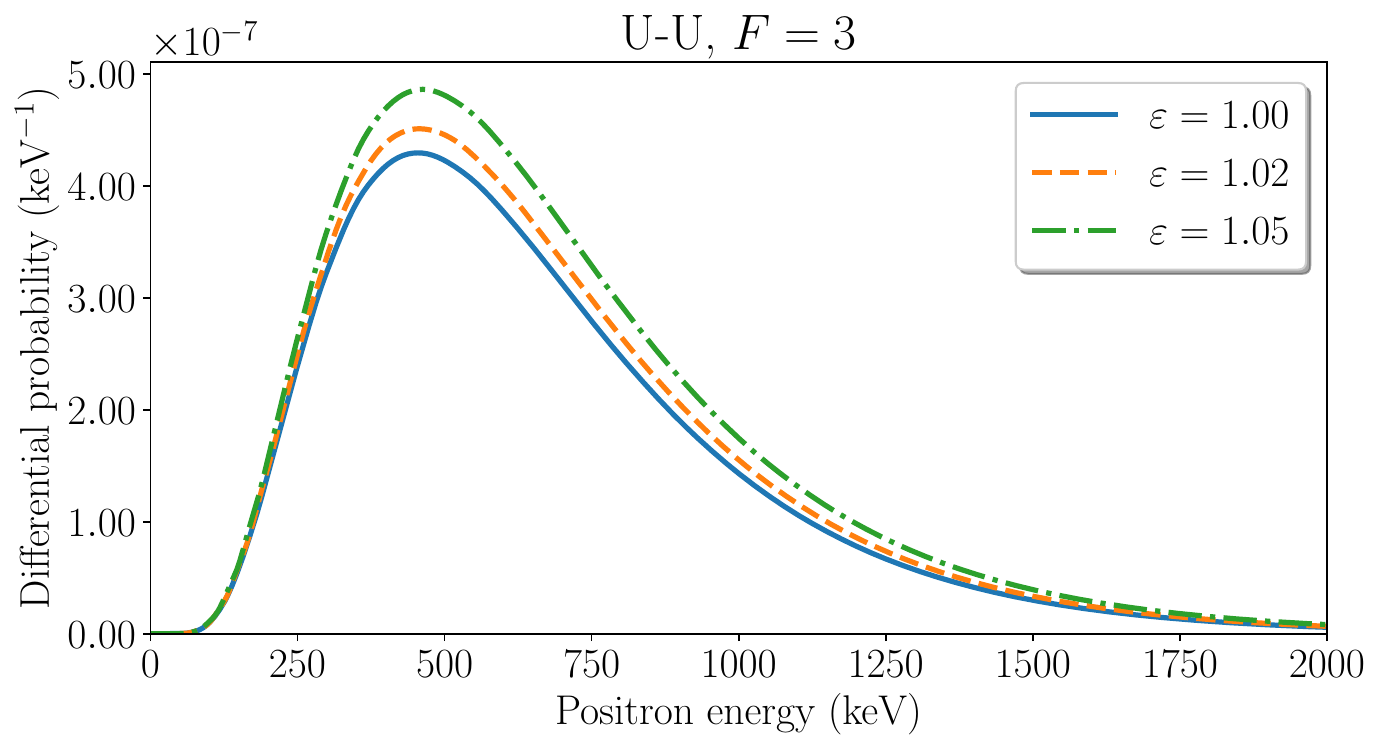}
    \caption{Energy spectra of positrons for symmetric collisions of nuclei with $Z=92$ and $F=3$ at $R_{\mathrm{min}}=17.5$~fm and $\varepsilon=1.0$, $1.02$, $1.05$ with included electron-electron interaction.}
    \label{fig:z_92_F_3_scr}
\end{figure}
\begin{figure}
    \centering
    \includegraphics[width=\columnwidth]{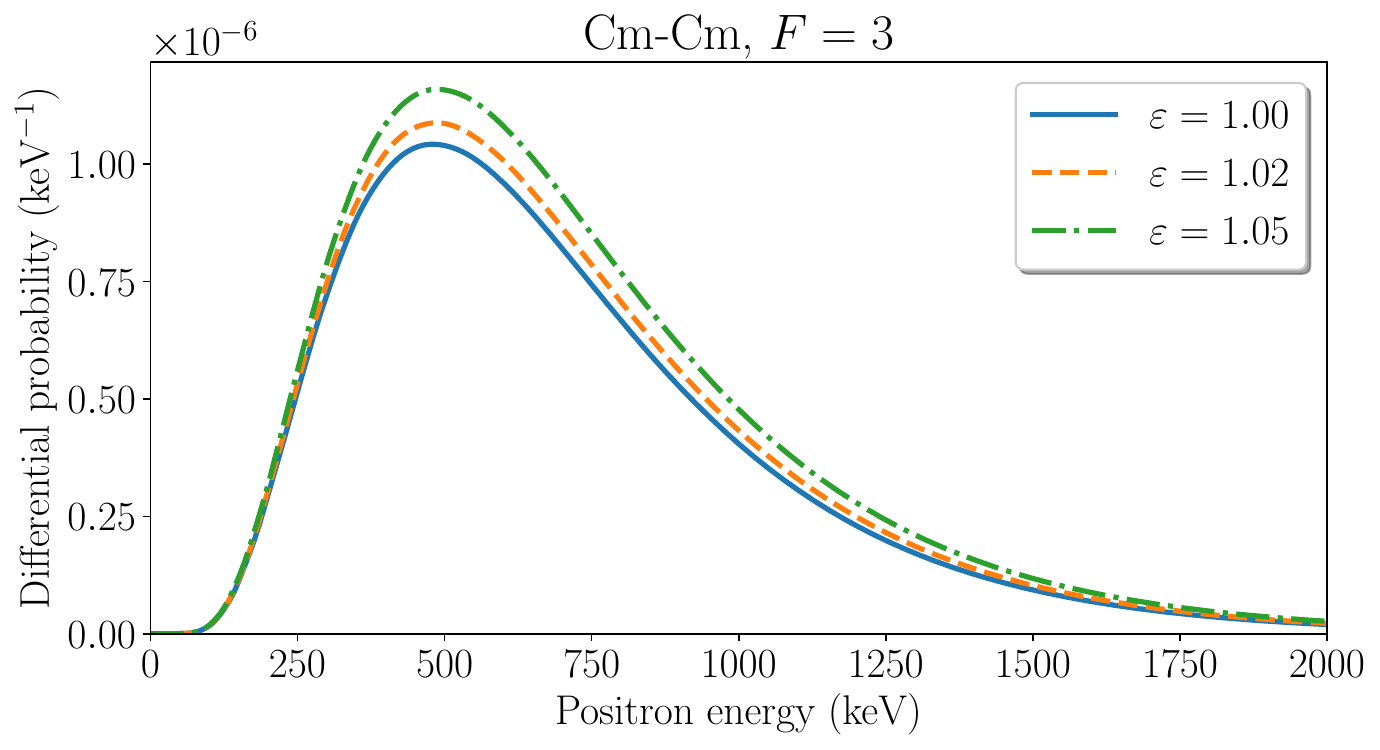}
    \caption{Energy spectra of positrons for symmetric collisions of nuclei with $Z=96$ and $F=3$ at $R_{\mathrm{min}}=17.5$~fm and $\varepsilon=1.0$, $1.02$, $1.05$ with included electron-electron interaction.}
    \label{fig:z_96_F_3_scr}
\end{figure}

Figures \ref{fig:z_92_F_3_scr} and \ref{fig:z_96_F_3_scr} present the positron energy distributions for U–U and Cm–Cm collisions, respectively,  at $\varepsilon=1.0$, $1.02$ and $1.05$ with $R_{\mathrm{min}}=17.5$~fm. The spectra were obtained for the Fermi level $F=3$, including the electron-electron interaction. 
The spectra confirm the conclusions drawn from the total probability analysis: for both U–U and Cm–Cm collisions, the peaks of the energy distributions exhibit no signatures of the transition to the supercritical regime, growing higher with increasing $\varepsilon$.

\section{Conclusion}

In this work, we have examined the impact of the electron-electron interaction on positron production in supercritical heavy-ion collisions. Within the monopole approximation, we calculated total positron creation probabilities and energy distributions for U–U, U-Cm and Cm–Cm collision systems, considering both bare nuclei and ions with electronic states occupied up to the Fermi level $F=3$. The electron-electron interaction was described using the TDDFT and TDHFS theoretical methods. The agreement of our results with the previous studies confirms the reliability of the employed computational approaches.

Our analysis demonstrates that the presence of occupied electronic states leads to a substantial suppression of positron production, as the fully occupied lowest bound states cannot be filled with electrons created from the vacuum and thus no longer contribute to the positron production. Accordingly, the signatures of the spontaneous positron creation regime observed in collisions of bare nuclei disappear for $F=3$, and positron emission becomes predominantly driven by the dynamical rather than spontaneous mechanism. Accounting for the electron-electron interaction in the calculations leads to  further $\approx 20$\% decrease of the positron creation probabilities while the qualitative picture of the pair creation remains unchanged.

Our results show that the characteristic signatures of the transition to the supercritical regime, previously found in collisions of bare nuclei \cite{Maltsev_2019_How, Popov_2020_How, Popov_2023_Spontaneous, Dulaev_2024_Angular, Dulaev_Three_2025}, are strongly suppressed in collisions of partially stripped ions. 

\begin{acknowledgments}
The calculations within the TDDFT approach were supported by the Russian Science Foundation (Grant No 22-62-00004, \cite{rscf}). The calculations using the TDHFS method were supported by the Theoretical Physics and Mathematics Advancement Foundation ``BASIS'' (Grant No 23-1-1-54-3). Computations were held on the basis of the HybriLIT heterogeneous computing platform (LIT, JINR).
\end{acknowledgments}

\bibliography{main.bib}
\end{document}